\documentclass[sigconf]{acmart}
\usepackage{CJKutf8} % 中文支持
\usepackage{enumitem}
\usepackage{xcolor}

\AtBeginDocument{%
  }

\setcopyright{acmlicensed}
\copyrightyear{2018}
\acmYear{2018}
\acmDOI{XXXXXXX.XXXXXXX}

\acmConference[Conference acronym 'XX]{Make sure to enter the correct
  conference title from your rights confirmation emai}{June 03--05,
  2018}{XXXX, xx}

\acmBooktitle{XXXX 'xx: ACM,
 June 03--05, 2018, xxxx, xx}
\acmISBN{978-1-4503-XXXX-X/18/06}

\begin{document}
\begin{CJK*}{UTF8}{gbsn} % 中文支持
% \title{ETimeline: A Large-Scale Event-Centric Timeline Generation Dataset based on Large Language Model}
\title{ETimeline: An Extensive Timeline Generation Dataset based on Large Language Model}
% \title{A Large-Scale Event-Centric Timeline Dataset: Unraveling the Fabric of Chronological Narratives}
% \author{Yanan Zhang}
% \authornote{Equal Contributions.}
% \author{Xiaochen Liu}
% \affiliation{%
%   \institution{Tencent Seach}
%   \city{Bejing}
% \email{{yananzhang, ashxcliu}@tencent.com}
% % \email{ashxcliu@tencent.com}
\author{Xiaochen Liu}
\authornotemark[1]
\affiliation{%
  \institution{Tencent}
  \city{Bejing}
  \country{China}
}
\email{ashxcliu@tencent.com}

\author{Yanan Zhang}
\authornote{Equal Contributions.}
\affiliation{%
  \institution{Tencent}
  \city{Bejing}
  \country{China}}
\email{yananzhang@tencent.com}


\begin{abstract}
Timeline generation is of great significance for a comprehensive understanding of the development of events over time. Its goal is to organize news chronologically, which helps to identify patterns and trends that may be obscured when viewing news in isolation, making it easier to track the development of stories and understand the interrelationships between key events.
Timelines are now common in various commercial products, but academic research in this area is notably scarce. Additionally, the current datasets are in need of refinement for enhanced utility and expanded coverage.
In this paper, we propose \textbf{ETimeline}, which encompasses over $13,000$ news articles, spanning $600$ bilingual timelines across $28$ news domains. Specifically, we gather a candidate pool of more than $120,000$ news articles and employ the large language model~(LLM) Pipeline to improve performance, ultimately yielding the \textbf{ETimeline}. The data analysis underscores the appeal of \textbf{ETimeline}. Additionally, we also provide the news pool data for further research and analysis.
This work contributes to the advancement of timeline generation research and supports a wide range of tasks, including topic generation and event relationships. We believe that this dataset will serve as a catalyst for innovative research and bridge the gap between academia and industry in understanding the practical application of technology services. The dataset is available at \url{https://zenodo.org/records/11392212}.
\end{abstract}





\maketitle

\section{Introduction}
\begin{figure}[t]
  \centering
  \includegraphics[width=0.49\textwidth]{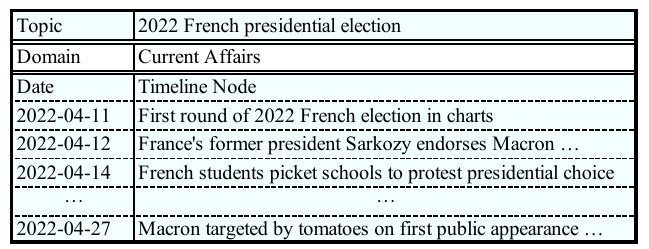} 
  % \vspace{-2em}
  \caption{Excerpt of the 2022 French presidential election timeline, using the news titles as nodes.} 
  % \vspace{-1em}
  \label{fig:timeline_sample} 
\end{figure}

Timeline generation helps us to comprehensively understand the development of events from a macro perspective. Its core value lies in arranging news events in chronological order to reveal deeper patterns and trends that may be overlooked when observed in isolation.
% With a timeline, we can more easily track the trajectory of each event, understand the intrinsic connections and mutual influences between different sub-events. It is s like the skeleton of a story, allowing us to clearly see each key node, and how they are interconnected, collectively driving the entire event forward.
With a timeline, we can more easily track the trajectory of each event, understand the intrinsic connections between different sub-events, their mutual influences, and how they collectively drive the entire event forward.
% For example, stock practitioners often need to make rational decisions from the development of international situations, such as the Israel-Palestine conflict, to changes in the price of pigs.
For example, stock market professionals often need to make informed decisions based on the evolution of global events, ranging from geopolitical conflicts like the Israel-Palestine situation to fluctuations in commodity prices such as pork.
Figure~\ref{fig:timeline_sample} shows the timeline of the 2022 French presidential election.

% \vspace{1em}
% \vspace{-0.5em}

Timelines are featured in many commercial products, such as search engines that display a timeline when you enter a query about an event, yet there is a noticeable lack of research in this field within academia. There is a significant gap between academia and industry in terms of research in this specific area.
Existing public datasets for timeline generation have the following shortcomings:
\begin{enumerate}[label=(\arabic*), leftmargin=*]
\item The data scale is small and lacks diversity. $17$ Timelines~(T17)~\citep{10.1145/2487788.2487829} and CRISIS~\citep{tran2015timeline} cover $17$ timelines in $9$ domains and $4$ timelines in $1$ domain respectively, while \citet{holt-etal-2016-presenting} proposes data containing $39$ entities. 
\item Previous methods are comparatively weak, making the data less difficult. These have mainly focused on unsupervised techniques\citep{holt-etal-2016-presenting, steen-, gholipour}, small-scale machine learning models~\citep{10.1145/2487788.2487829, tran2013leveraging, tran2015timeline, chen2019learning}, etc. However, the reality is that many pieces of information need robust inferential capabilities for association. For instance, events such as ``Trump potentially facing a criminal trial'' and ``Trump possibly being unable to run for president'' require causal reasoning to determine their connection, beyond mere semantic similarity. This level of complexity often makes previous NLP methods insufficient for the task.
\end{enumerate}

To this end, we introduce \textbf{ETimeline}, which features $600$ bilingual timelines and 13,878 event nodes. These span a range of popular Internet events from March $2020$ to April $2024$, across $28$ news domains~($\S$~\ref{sec:data_a}). Additionally, the original news pool utilized for timeline generation is also included.
We also propose a pipeline based on LLM for data construction. This pipeline gathers popular topics and news reports from the Internet, simulating a streaming timeline generation process. Utilizing LLMs, significantly enhances the performance of timeline generation, establishing a robust baseline that helps bridge the substantial gap between academia and industry in the field~($\S$~\ref{sec:data_c}).

% \textbf{ETimeline} is a large-scale, diverse, and high-performance dataset that provides a new and competitive option for the timeline generation task. In addition, it also supports a wide range of tasks, including~(multi-level) topic generation, event relationships, and more.
\textbf{ETimeline} is a large-scale, diverse, and high-performing dataset that offers a novel and competitive alternative for timeline generation tasks. Additionally, it supports a broad spectrum of tasks, including topic generation, event relationship analysis, and more.
% \section{maybe need related work}
% exist dataset ?llm?
% 待定

\section{DATA CONSTRUCTION}\label{sec:data_c}

We source data from the Internet and stream it to construct \textbf{ETimeline}, which mirrors real-world processing logic. Additionally, we meticulously craft a pipeline using a $7$B-parameter LLM to enhance dataset performance.
Specifically, the pipeline is generalized into two phases: 
\begin{enumerate}[label=(\arabic*), leftmargin=*]

 \item \textbf{Timeline topic extraction}. We gather trending words from platforms like \textit{Google Trends} and \textit{Baidu Hotsearch}. The LLM then crafts appropriate topics from these trends, refining them as new nodes are integrated into the timeline. 

 \item \textbf{Timeline node filling}. We assemble a chronologically arranged pool of potential news documents, culled from online news sources.  The LLM then methodically designates each document to a node under the appropriate topic, emulating the real-world continuous emergence and processing of news by the pipeline.
\end{enumerate}

% \vspace{-0.5em}
\begin{figure}[t]
  % \centering
  \hspace*{-\parindent} \includegraphics[width=0.48\textwidth]{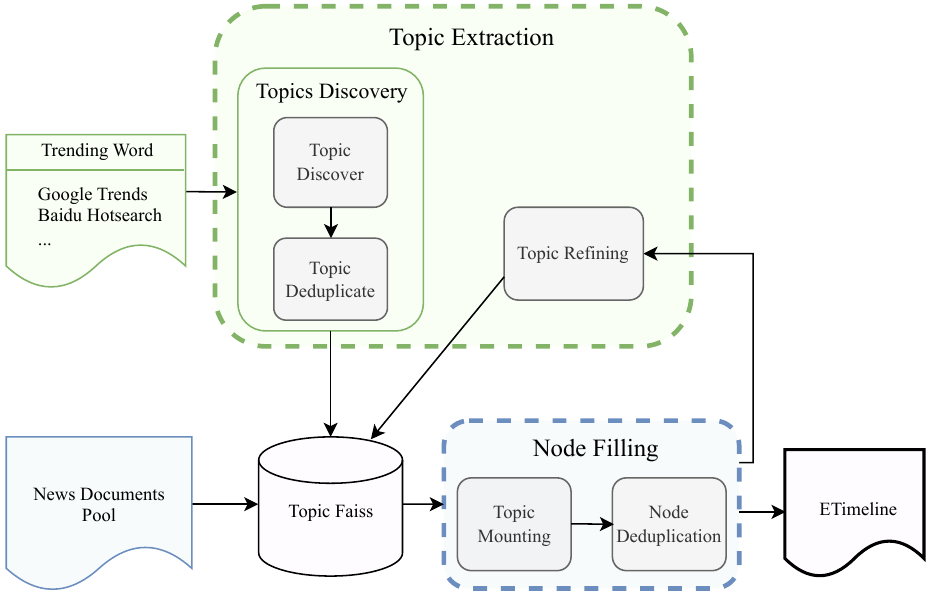} 
  % \vspace{-1em}
  \caption{Data construction pipeline for ETimeline.} 
  % \vspace{-1em}
  \label{fig:pipline_pic} 
\end{figure}
 % \vspace{-0.5em}
 
Integral to this process, the LLM undertakes four key functions: topic generation, topic refining, topic mounting, and text deduplication. To optimize the capabilities of the $7$B-parameter LLM, we employ knowledge distillation techniques, utilizing GPT-$4$~\cite{achiam2023gpt} for initial data annotation, followed by the training of a smaller, more readily deployable model for data construction.
The overall pipeline is illustrated in Figure~\ref{fig:pipline_pic}.

% \vspace{-0.5em}
\definecolor{myblue}{RGB}{102,255,255} 
\begin{figure}[t]
  % \centering
  \includegraphics[width=0.5\textwidth]{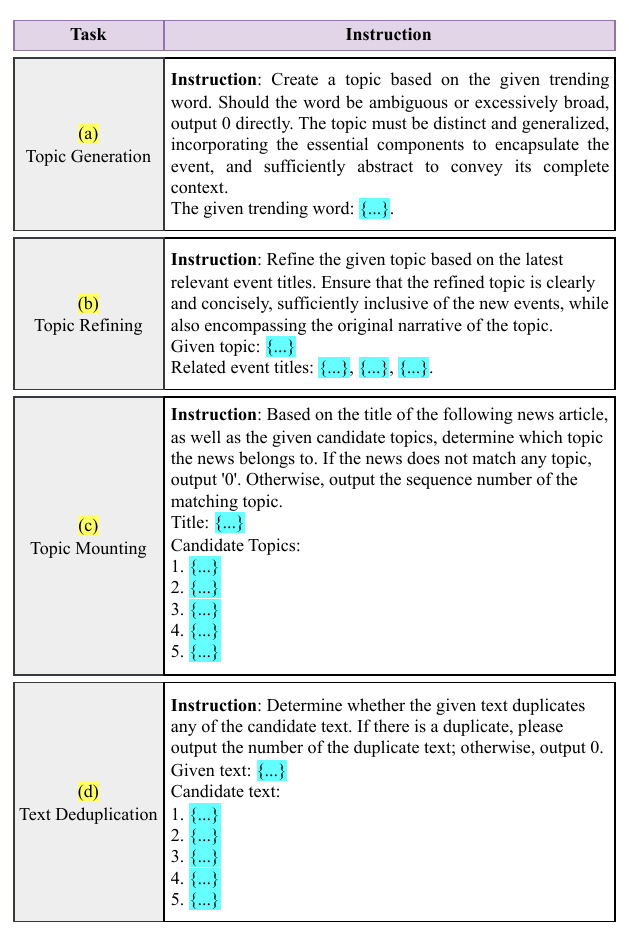} 
  % \vspace{-1em}
  \caption{Prompts for different tasks executed by LLM in the data construction pipeline. The text highlighted in \colorbox{myblue}{light blue} would change dynamically with different inputs.} 
  % \vspace{-1em}
  \label{fig:prompt_pic} 
\end{figure}
 % \vspace{-0.5em}
\subsection{Timeline Topic Extraction}
\subsubsection{Topic Discovery}
Topic discovery initiates with the analysis of trending terms sourced from platforms like \textit{Google Trends} and \textit{Baidu Hotsearch}, which swiftly and broadly encompass significant events.
Firstly, we utilize the LLM to parse these trending words for topic generation, encompassing two key facets:
\setlist[itemize]{leftmargin=*}

\begin{itemize}
\item \textbf{Low-quality word filtering:} We eliminate terms that are ambiguously articulated and overly broad.
\item \textbf{Topic formulation:} From the remaining keywords, we craft topics intended to be distinct, succinct, and encapsulating.
\end{itemize}
These dual considerations are amalgamated into a unified prompt for execution, depicted in Figure~\ref{fig:prompt_pic}(a).

Given the potential for duplicate topics, such as identical events captured by different sources' trending words or the emergence of new trending terms as an event unfolds, we integrate a deduplication process.
The newly minted topics will seek out the top-$k$ most similar ones within the existing topics' faiss~\cite{johnson2019billion} vector index, and subsequently evaluated by the LLM for redundancy. Duplicates are culled, while unique topics are integrated into the topic faiss library. Figure~\ref{fig:prompt_pic}(d) illustrates the prompt employed to ascertain topic duplication via the LLM.

\subsubsection{Topic Refining} 

The ever-changing tapestry of real-world events demands ongoing updates and refinement of the topics. For example, an initial topic like ``Russian troops mass on Ukraine's border'' might evolve into ``Russia-Ukraine war'' as new events such as ``Diplomatic Talks Initiated'', ``Sanctions Imposed'', and ``Military Clashes Reported'' unfold. In our pipeline, the addition of each new event to the timeline prompts a refresh of the associated topics.
 
With the guidance of the prompt depicted in Figure~\ref{fig:prompt_pic}(b), we make extensive use of an LLM to ensure that our topics encapsulate the current timeline in a way that is both thorough and succinct:
\setlist[itemize]{leftmargin=*}

\begin{itemize}
\item \textbf{Incorporating fresh insights:} We craft updated topics by incorporating the existing topic and the titles of the most recent $k$ nodes.
\item \textbf{Elevating abstraction:} As new events are integrated, the LLM refines the topic to capture a broader perspective.
\end{itemize}
% The dynamic nature of real-world events necessitat
\subsection{Timeline Node Filling}
The timeline nodes originate from news documents collected from the Internet, spanning from March 2020 to April 2024. To ensure quality and reliability, we apply several heuristic filtering rules to preprocess the documents, specifically targeting low-quality pages, non-authoritative sources, and dead links. The remaining documents are organized into a pool arranged in chronological order to simulate the continuous occurrence of events in the real world.

The timeline node filling process includes:~(1) attaching new event nodes to appropriate topics, and~(2) deduplicating the new nodes with existing nodes on the timeline and inserting them in the correct position.

\subsubsection{Topic Mounting} 
The process of topic mounting kicks off by leveraging the titles of streaming documents to query the topic faiss library, which was established during the phase of topic extraction. This preliminary search step narrows down a selection of potential topics that could correspond to the current article.

Upon assembling a list of potential topics, we craft a prompt that integrates the title of the current article along with the identified candidate topics, as illustrated in Figure~\ref{fig:prompt_pic}(c). This prompt is then presented to the LLM. The LLM assesses the prompt, discerning if the article qualifies to be appended as an event node under one of the candidate topics. If it does, the LLM further selects the most fitting topic for mounting.

\subsubsection{Node Deduplication}
After the topic mounting phase, articles associated with the candidate topics are integrated as event nodes beneath those topics. Considering the prevalence of multiple similar reports on the same event, a deduplication process is mandatory for every new document prior to its inclusion in the timeline.
We initially employ straightforward textual similarity measures, such as BM25~\citep{bm25-2009} and n-gram matching ratio, to pinpoint the top-$k$ most analogous potential duplicates on the timeline that is pertinent to the current article.

Subsequently, we utilize once more the prompt illustrated in Figure~\ref{fig:prompt_pic}(d) to encompass both the current document and the recognized candidate nodes. Through the LLM, we evaluate whether the new entry matches any existing nodes as a duplicate.
Should a duplicate be detected, the current document is rejected to avoid redundancy; if not, it is seamlessly positioned into the timeline in its proper chronological sequence.

\begin{table}[t]
    \caption{Dataset statistics of ETimeline.}
    \begin{tabular}{lc}
    \toprule
    \textbf{Statistic}            & \textbf{Number} \\ \midrule
    Total timelines               & 600            \\
    Total nodes                  & 13,878             \\
    Total news domain     & 28             \\
    Language     & en/zh             \\
    News pool capacity & 127,170           \\ \bottomrule
    \end{tabular}
    % \vspace{8pt}

    \label{tab:stats}
\end{table}

\subsection{Distillation}
We harness the power of a $7$B-parameter large language model\footnote{For English data, we adopt the Mistral-7B-Instruct model~(\url{https://huggingface.co/mistralai/Mistral-7B-Instruct-v0.2}); for Chinese data, we select the Baichuan2-7B-Chat model~(\url{https://huggingface.co/baichuan-inc/Baichuan2-7B-Chat}).} to execute the full data construction pipeline, which includes four pivotal tasks: topic generation, topic refining, topic mounting, and text deduplication.
% Considering factors such as inference latency and cost, a model with this magnitude of parameters is more suitable for industrial deployment.
These models, with their parameter size, are better suited for real-world applications, considering factors like inference speed and cost-effectiveness.
To further boost the capabilities of the $7$B LLM, we integrate a knowledge distillation strategy.

For these four tasks, we meticulously select data and crafted task-specific directives to extract annotations from GPT-$4$, encompassing both answers and rationales. After applying heuristic rules for filtering, we assemble a curated dataset of $10,000$ training examples per task, amassing a total of $40,000$ examples.
The $7$B-parameter model underwent supervised fine-tuning using these examples. $50\%$ of the training dataset encompassed annotated answers along with their reasons, designed to bolster the reasoning abilities of the model. The other half of the dataset focused exclusively on answers, with the goal of solidifying the model's output formatting. The process of model distillation has significantly improved the performance of the $7$B LLM.

\section{DATA ANALYSIS}\label{sec:data_a}
% \vspace{-1em}
\begin{figure}[t]
  \centering
  \includegraphics[width=0.45\textwidth]{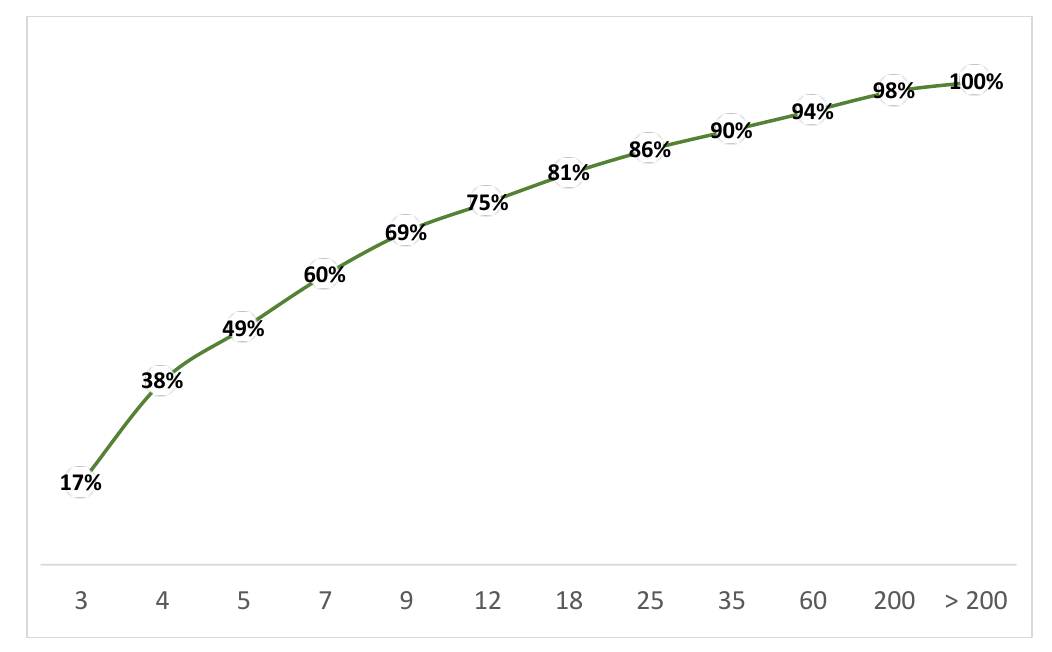} 
  % \vspace{0em}
  \caption{Cumulative distribution of the timeline lengths. The x-axis displays the timeline lengths, while the y-axis illustrates the cumulative percentage.} 
  % \vspace{-1em}
  \label{fig:node_s} 
\end{figure}
% \vspace{-0.5em}
\subsection{Statistics}
\begin{figure}[t]
  \centering
  \includegraphics[width=0.45\textwidth]{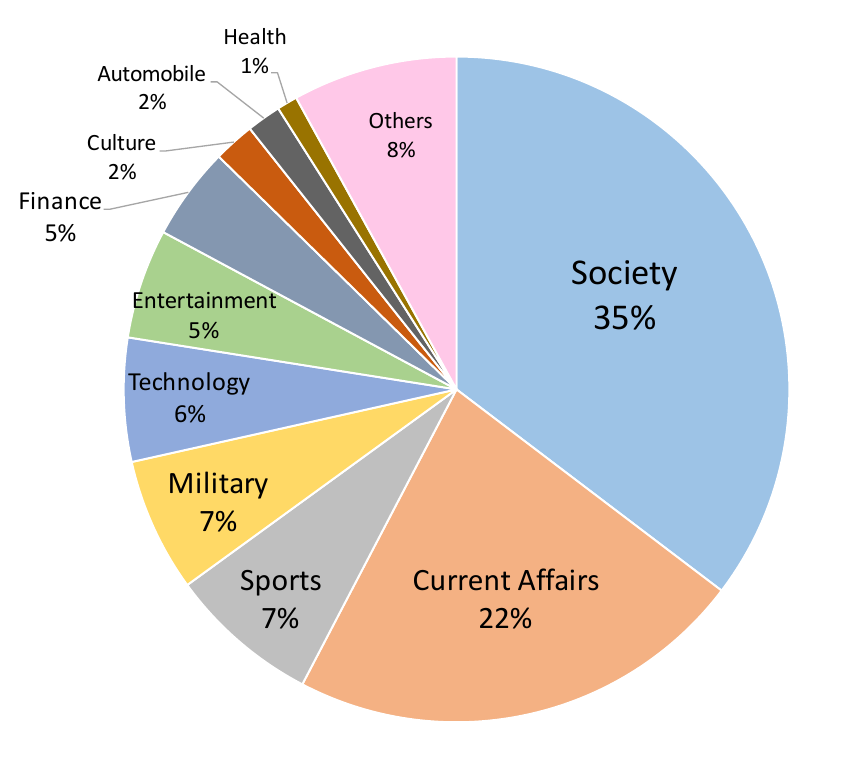} 
  % \vspace{-1em}
  \caption{The distribution of news domain in ETimeline, all non top-10 topics are aggregated as ``Others''.} 
  % \vspace{-1em}
  \label{fig:domain_s} 
\end{figure}

\textbf{Total Scale.}\quad
Table~\ref{tab:stats} shows the overall situation of \textbf{ETimeline}, with the data scale being notably prominent.
\noindent \\
\textbf{Timeline Length.}\quad
Figure~\ref{fig:node_s} displays the cumulative distribution of the timeline lengths. Timelines with more than $10$ nodes account for $30\%$, posing a considerable challenge.
It should be noted that we only preserve those timelines that contain $3$ or more nodes, based on the rationale that an event lifecycle should minimally include three stages: initiation, progression, and resolution.
\noindent \\
\textbf{Domain Distribution.}\quad
The dataset spans $28$ news domains, ensuring sufficient diversity.
Figure~\ref{fig:domain_s} shows the distribution of the top-$10$ domains to which the timelines belong. 
% \vspace{-1em}
% \vspace{-0.5em}

\subsection{Striking Data}
% \noindent \\
\textbf{Topic Diversity.}\quad
We did not deliberately standardize the topic formats, such as requiring them to follow the composition of events - necessarily including triggers and arguments. Therefore, there are some special cases where topics are a single entity, such as ``Israel-Palestine conflict'', which is sufficiently inclusive and concise.
\noindent \\
\textbf{Data Multifacetedness.}\quad
We objectively treat each collected news document, respecting its expressed emotional attitude, and place it in its corresponding timeline. As a result, various positive or negative nodes can be seen within the same timeline, such as the positive and negative news about Macron's candidacy for the French presidency in Table~\ref{fig:timeline_sample}.

\section{POTENTIAL APPLICATIONS}
By releasing \textbf{ETimeline}, we aim to foster the timeline generation task.
Researchers can use the provided news pool to refine the timelines.
We also hope our innovative data construction method will draw more researchers to this field, as previous NLP methods are inadequate for its complexity and the industry has already gone further.

Taking into account the intricate nature of the timeline generation dataset, brimming with a multitude of events, \textbf{ETimeline} presents opportunities for several other auspicious research trajectories:

\setlist[itemize]{leftmargin=*}
\begin{itemize}
\item \textbf{Event Relationship Modeling.}
Understanding the interconnections among events is crucial for mapping out their developmental progression. Frequently, distinguishing the relationship between two events necessitates intricate reasoning and analysis. By integrating the timeline's node data with more robust NLP models, such as LLMs, one can significantly bolster the efficacy of event relationship modeling.

\item \textbf{Topic Generation.} 
Inclusive and concise topics facilitate a rapid understanding of events. Generating themes from individual or multiple news documents is challenging. Moreover, multi-level topic analysis is sometimes required, such as ``Israel-Palestine conflict'', which can be further divided into the actions of Israel, Palestine, and other countries. This can aid politicians, historians, and others in gaining a more comprehensive understanding from various viewpoints.
\end{itemize}

\section{CONCLUSIONS}
We release \textbf{ETimeline}, an extensive timeline generation dataset based on large language models, which includes bilingual Chinese and English, $13,000$ nodes, $600$ timelines, covering popular events that occurred from April $2020$ to April $2024$. We detail the process of data construction, propose a robust baseline for timeline generation based on LLM that simulates real-world application scenarios, and suggest potential directions for future research. We hope that \textbf{ETimeline}, as a competitive candidate, will attract more research to bridge the huge gap between industry and academia in this task, and promote more event-related research.

\section{ETHICAL CONSIDERATIONS}
Our data collection fully comply with the platform's terms of service. Since the basic data of \textbf{ETimeline} is collected from the Internet, it is extremely difficult to remain unbiased. We have made great efforts to remove toxicity from the data, preprocess the news documents, and exclude potential pornography and personal information. However, we have retained objective positive or negative declarative documents and believe they are positively meaningful for a comprehensive understanding of a timeline. Instead of sharing original content, we offer titles and URLs for accessing the data, which can be used to retrieve original data, except for deleted documents by the platform.

% %%
% %% The acknowledgments section is defined using the "acks" environment
% %% (and NOT an unnumbered section). This ensures the proper
% %% identification of the section in the article metadata, and the
% %% consistent spelling of the heading.
% \begin{acks}
% \end{acks}

\bibliographystyle{ACM-Reference-Format}
\bibliography{sample-base}

\appendix
\end{CJK*}
\end{document}